\documentclass[pra,twocolumn,showpacs]{revtex4}
\usepackage{graphicx}
\newcommand{\sq}
{\nobreak \ifvmode \relax \else \ifdim\lastskip<1.5em
\hskip-\lastskip \hskip0.25em plus0em minus0.0em \fi \nobreak \vrule
height0.75em width0.75em depth0.0em\fi}
\begin{document}

\def\C{{\rm I\!\!\!\! C}}

\title{Electromagnetic analysis of arbitrarily shaped pinched carpets}

\author{Guillaume Dupont,$^{*}$ S\'ebastien Guenneau,$^{*}$
and Stefan Enoch$^{*}$}
\affiliation{$^*$Institut Fresnel, CNRS,
Aix-Marseille Universit\'e,
\\
Campus Universitaire de Saint-J\'er\^ome, 13013 Marseille, France
\\}
\date{\today}
\begin{abstract}
We derive the expressions for the anisotropic heterogeneous tensors
of permittivity and permeability associated with two-dimensional and
three-dimensional carpets of an arbitrary shape. In the former case,
we map a segment onto smooth curves whereas in the latter case we
map a non convex region of the plane onto smooth surfaces.
Importantly, these carpets display no singularity of the
permeability and permeability tensor components, and this may lead
to some broadband cloaking.
\end{abstract}
%
\maketitle 

\def\R{{\rm I\!R}}
\newcommand{\beq}{\begin{eqnarray}}
\newcommand{\eeq}{\end{eqnarray}}

In 2006, the physicists Pendry, Schurig and Smith theorized that a
finite size object surrounded by a spherical coating consisting of a
metamaterial might become invisible for electromagnetic waves
\cite{pendrycloak}. This is somewhat analogous to the alternative
route to invisibility using conformal mappings (in the complex
plane), preferred by Leonhardt \cite{leonhardt}. These two proposals
have captured the imagination of scientists working in the area of
metamaterials. However, the former is not restricted by small
wavelengths, and it has been experimentally validated in the
microwave regime using a two-dimensional setup \cite{cloakex}.

The underlying idea behind the cloaking using transformation optics
is to map a point in optical space onto a spherical (invisibility)
region. Back in 1984, the mathematicians Kohn and Vogelius noticed
that one could find the conductivity of an object from static
measurements on its boundary \cite{kohn84}. In the same vein,
Greenleaf, Lassas and Uhlmann looked in 2003 at an inverse problem
where the Dirichlet to Neumann map defining a coating had the
required properties to make a small conducting body nearly invisible
\cite{greenleaf}. But the important physical consequences had not be
drawn by the mathematicians.

Many authors have since then dedicated a fast growing amount of work
to the invisibility cloaking problem. Interestingly, there are
alternative approaches,  including some which make use of plasmonic
properties of coated cylinders \cite{milton2,engheta}. These latter
proposals are sometimes referred to as external cloaking. The main
advantage over the transformation optics approach is that there is
no requirement for anisotropic heterogeneous permittivity and
permeability, which is a consequence of the change of coordinates
\cite{ward,nicolet,Ulf,post}. However, external cloaking is
narrowband in nature, whereas transformation optics allows for
broadband cloaking, and works even in the intense near field limit
when a source is located a couple of wavelengths away from the cloak
\cite{zgnp07}. Transformation optics can also be used to design
generalized perfect lenses \cite{pendry_jpc03}.


A severe limitation in the design of invisibility cloaks via
transformation optics is the singular behaviour of the material
parameters at the cloaks' inner boundary, which is a consequence of
tearing apart the metric when one makes a hole in optical space
(known in mathematics as blow up theory \cite{greenleaf}).
Physically, light has to curve its trajectory around the hole (or
'invisibility region'); Hence, to match the phase of a wave
propagating in homogeneous space, it must travel faster. One way to
avoid such paradoxes is to approximate the cloaks' parameters using
a homogenization approach, which leads to nearly ideal cloaking
\cite{naturecloak,farhat,kurylev}. Attractive theoretical proposals
to avoid the cloaks' singularities include the design of nearly
ideal (non-singular) two dimensional cloaks from a projection of
three dimensional ideal (but singular) cloaks
\cite{heapl,zhangcarpet,tyc}. An alternative route is to use a
one-to-one mapping to design an invisibility carpet, which is the
bottom line of the bold proposal by Li and Pendry to conceal an
object that is placed under a curved reflecting surface by imitating
the reflection of a flat surface \cite{lipendry}. The present letter
is the first report of arbitrarily shaped two-dimensional and
three-dimensional carpets.

In electromagnetism, a change of coordinates induced by a geometric
transform leads to the design of complex materials. For instance, if
we start from a homogeneous and isotropic dielectric medium
described by a permittivity $\varepsilon$ and a permeability $\mu=1$
(no magnetism), we end up with an inhomogeneous anisotropic material
described by a transformation matrix ${\bf T}$ (also known as metric
tensor) \cite{ward,nicolet,zgnp07,Ulf}. The permittivity and
permeability in the transformed coordinates are now given by:
\begin{equation}
\underline{\underline{\varepsilon'}} =\varepsilon \mathbf{T}^{-1} \;
,  \quad
 \hbox{and} \quad
\underline{\underline{\mu'}}=\mu \mathbf{T}^{-1} \hbox{ where
$\mathbf{T} \!= \! \mathbf{J}^T \mathbf{J}/det(\mathbf{J})$} \; ,
\label{epsmuT}
\end{equation}
where $\mathbf{J}$ is the Jacobian matrix of the transformation.
Importantly, we note that this material is magnetic.

We now want to apply this recipe to design two-dimensional and
three-dimensional carpets.
Let us first consider the linear geometric transform:
\begin{equation}
\left\{
\begin{array}{ll}
x' &= x \; , \; a< x < b \; ,\\
y' &= \frac{y_2(x)-y_1(x)}{y_2(x)}y + y_1(x) \; , \; 0<y< y_2(x) \; ,\\
z' &= z \; , \; -\infty< z < +\infty \; ,
\end{array}
\right. \label{transfo2D}
\end{equation}
where $y'$ is a stretched vertical coordinate. It is easily seen
that this linear geometric transform maps the segment $(a,b)$ of the
horizontal axis $y=0$ onto the curve $y'=y_1(x)$, and it leaves the
curve $y=y_2(x)$ unchanged. Importantly, there is a one-to-one
correspondence between the segment and $y_1$. The curves $y_1$ and
$y_2$ are assumed to be differentiable, and this ensures that the
carpet won't display any singularity on its inner boundary, as we
shall now derive.

The linear transform (\ref{transfo2D}) is expressed in a Cartesian
basis as: ${\bf J}_{xx'}=\left(
\begin{array}{ccc}
                1 & 0 & 0\\
             \frac{\partial y}{\partial x'} & \frac{1}{\alpha} & 0\\
             0 & 0 & 1\\
               \end{array}\right)$
where $\alpha=(y_2-y_1)/y_1$ and from the chain rule
\begin{equation}
\frac{\partial y}{\partial x'}= y_2 \frac{y'-y_2}{{(y_2-y_1)}^2}
\frac{\partial y_1}{\partial x} - y_1 \frac{y'-y_1}{{(y_2-y_1)}^2}
\frac{\partial y_2}{\partial x} \; .
\end{equation}

This leads to the inverse symmetric tensor ${\bf T}^{-1}$ which is
fully described by five non vanishing entries in a Cartesian basis:
\begin{equation}
\begin{array}{ll}
(T^{-1})_{11}=\displaystyle \frac{1}{\alpha} \; ,
(T^{-1})_{12}=(T^{-1})_{21}=-\displaystyle \frac{\partial
y}{\partial x'}\\
(T^{-1})_{22}=\displaystyle \left( 1+{\left(\frac{\partial
y}{\partial x'}\right)}^2\right) \alpha \; ,
(T^{-1})_{33}=\displaystyle \frac{1}{\alpha}
\end{array}
\label{invt2d}
\end{equation}

It is interesting to look at the behaviour of the eigenvalues of
${\bf T}^{-1}$ as these are the relevant quantities to compute the
tensor components along the main optical axes:
\begin{equation}
\begin{array}{ll}
\lambda_1=\displaystyle{\frac{1}{\alpha}},
\lambda_{i}=\displaystyle{\frac{1}{2\alpha}\left(1+\alpha^2+\left(
\frac{\partial
y}{\partial x'} \right)^2 \alpha^2 \right.} \\
\left. + (-1)^{i-1}\sqrt{-4\alpha^2+ \left(1+\alpha^2+\left(
\frac{\partial y}{\partial x'} \right)^2 \alpha^2 \right)^2}\right).
\end{array}
\label{lambda2d}
\end{equation}

We note that $\lambda_1$ and $\lambda_i$, $i=2,3$, are strictly
positive functions as obviously $1+\alpha^2+\left( \frac{\partial
y}{\partial x'} \right)^2 \alpha^2>\sqrt{-4\alpha^2+
\left(1+\alpha^2+\left( \frac{\partial y}{\partial x'} \right)^2
\alpha^2 \right)^2}$ and also $\alpha
>0$. This establishes that ${\bf T}^{-1}$ is not a singular matrix for a
two-dimensional carpet, which is a big advantage over
two-dimensional cloaks obtained by blowing up a point onto a disc
\cite{pendrycloak,greenleaf,zgnp07}: the transformation matrix is
then singular at the cloak's inner boundary (one eigenvalue goes to
infinity, while the other two go to zero).


For the sake of illustration, let us now consider a two-dimensional
carpet that has inner and outer boundaries given by
\begin{equation}
y_i(x)=h_i \left( e^{-\frac{1}{2}{(\frac{x}{\sigma})}^2}
-\frac{1}{8} \right) +c_i \sin\left(d_i.h_i \left(
e^{-\frac{1}{2}{(\frac{x}{\sigma})}^2}-\frac{1}{8}\right)\right) \;
,
\end{equation}
$i=1,2$, with $h_1=0.2$, $h_2=0.4$, $c_1=c_2=0.01$, $d_1=60$,
$d_2=50$ and $\sigma=0.3$.

\begin{figure}[h!]
\vspace{1cm}
\centerline{
\includegraphics[width=10cm,angle=0]{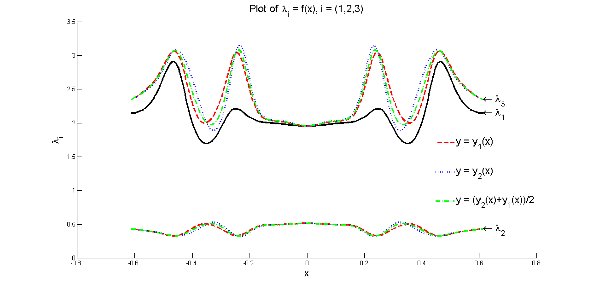}}
\vspace{1cm} \caption{Profile of the eigenvalues $\lambda_i(x)$,
$i=1,2,3$ of ${\bf T}^{-1}$ on the inner boundary $y_1(x)$, the
outer boundary $y_2(x)$ and the centerline $(y_2(x)+y_1(x))/2$ of
the carpet.} \label{fig0}
\end{figure}

We plot the profile of the three eigenvalues $\lambda_i$ along the
inner and outer boundaries $y_1$ and $y_2$ of the carpet, as well as
along the curve located half way from these, i.e.
$(y_2(x)+y_1(x))/2$. We can see in Fig. \ref{fig0} that none of the
eigenvalues vanish and they satisfy the inequality
$0<\lambda_2\leq\lambda_1\leq\lambda_3$, a fact which can be also
readily shown. We further note that $\lambda_1$ and $\lambda_2$ take
values strictly within $1.5$ and $3.5$.

Thanks to the cylindrical geometry, the problem splits into $p$ and
$s$ polarizations. In p polarization, we have:
\begin{equation}\label{eq:Hl}
\nabla\cdot\left( \underline{\underline{\varepsilon'_T}}^{-1}\nabla
{H}_3 \right) + \mu_0\varepsilon_0\omega^2 \mu'_{33}{H}_3= 0
\end{equation}
in the carpet, where ${\bf H}_l=(H_3(x,y)
-H_3^i(x,y)){\bf e}_z$ is the diffracted field parallel to the
cylinder axis. Importantly, ${\bf H}_l$ satisfies the usual outgoing
wave conditions as well as the Neumann data $\partial H_3/\partial
n=\partial H_3^i/\partial n$ on the ground plane and the inner
boundary of the cloak, with $H_3^i(x,y) {\bf e}_z$ the incident
field which is a beam generated by a constant field on a segment
located at the upper left corner of the computational domain, and
making an angle of $45$ degrees with the horizontal axis. Moreover,
$\underline{\underline{\varepsilon'_T}}$ is the upper left block
diagonal part of $ \underline{\underline{\varepsilon'}}$ and
$\mu'_{33}$ the third diagonal entry of
$\underline{\underline{\mu'}}$, as deduced from (\ref{epsmuT}) and
(\ref{invt2d}).

Such an anisotropic permittivity
$\underline{\underline{\varepsilon'_T}}$ could be achieved e.g.
using some thin wires of metal diluted in dielectrics
\cite{pendry96,ultrastef} to meet the condition that its eigenvalue
$\lambda_2$ is lower than $1$, see Fig. \ref{fig0}. Moreover,
$\mu'_{33}=\lambda_1$ involves some artificial magnetism which would
require some resonant elements such as split ring resonators
\cite{pendry_IEEE} used in the design of the first invisibility
cloak \cite{cloakex}.

\begin{figure}[h!]
\centerline{
\includegraphics[width=8cm,angle=0]{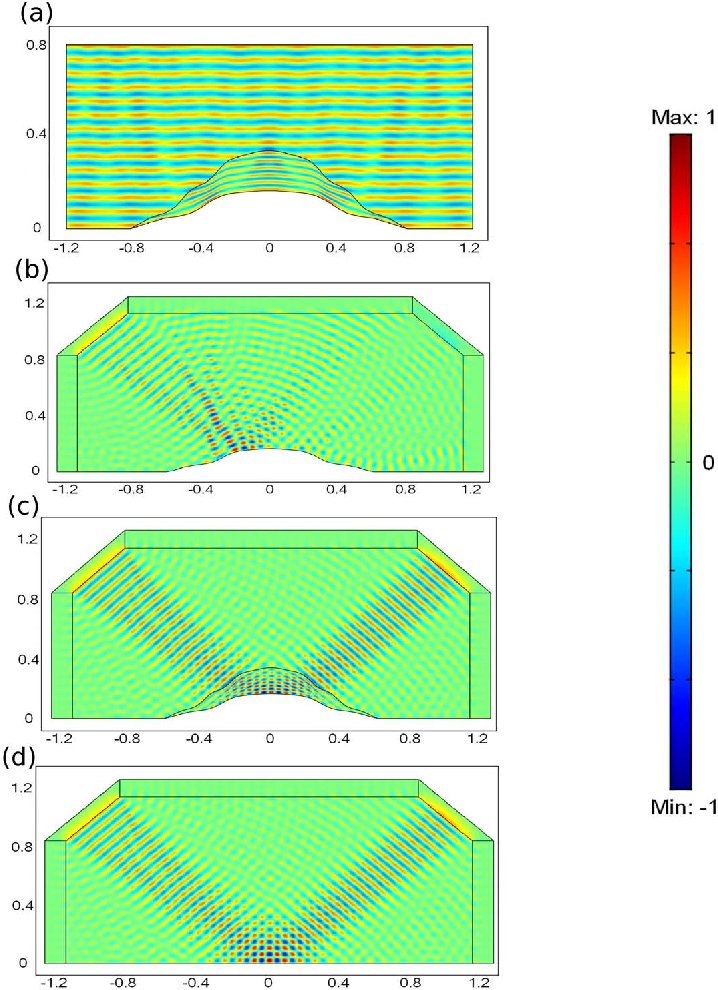}}
\caption{Diffraction by a plane wave and a beam at wavelength
$\lambda=0.06$: we set $H_3^i=1$ on the upper left side of the inner
trapezoidal domain; 2D plot of the real part of the component $H_3$
of the magnetic field.
(a) Deformed mirror with a carpet
under normal incidence; (b) Deformed mirror under oblique incidence;
(c) Same as (b) with a carpet; (d) Flat mirror under oblique
incidence.} \label{fig1}
\end{figure}

In what follows, we consider a plane wave and a beam incident from
the top at the wavelength $\lambda=0.06$. In Fig. \ref{fig1}, we
report some computations where the plane wave is coming from above
and the beam is incident from the top left corner, making an angle
of $\theta=45$ degrees with the normal to the ground plane.
We note that the field diffracted by the flat ground plane with
infinite conducting condition i.e. a mirror, cf. Fig. \ref{fig1}(d),
and by an infinite conducting object i.e. a curved mirror surrounded
by the carpet, cf. Fig. \ref{fig1}(c), indeed superimpose. Of
course, the field diffracted by the curved mirror on its own, cf.
Fig. \ref{fig1}(b), is much different.

\begin{figure}[h!]
\centerline{
\includegraphics[width=7cm,angle=0]{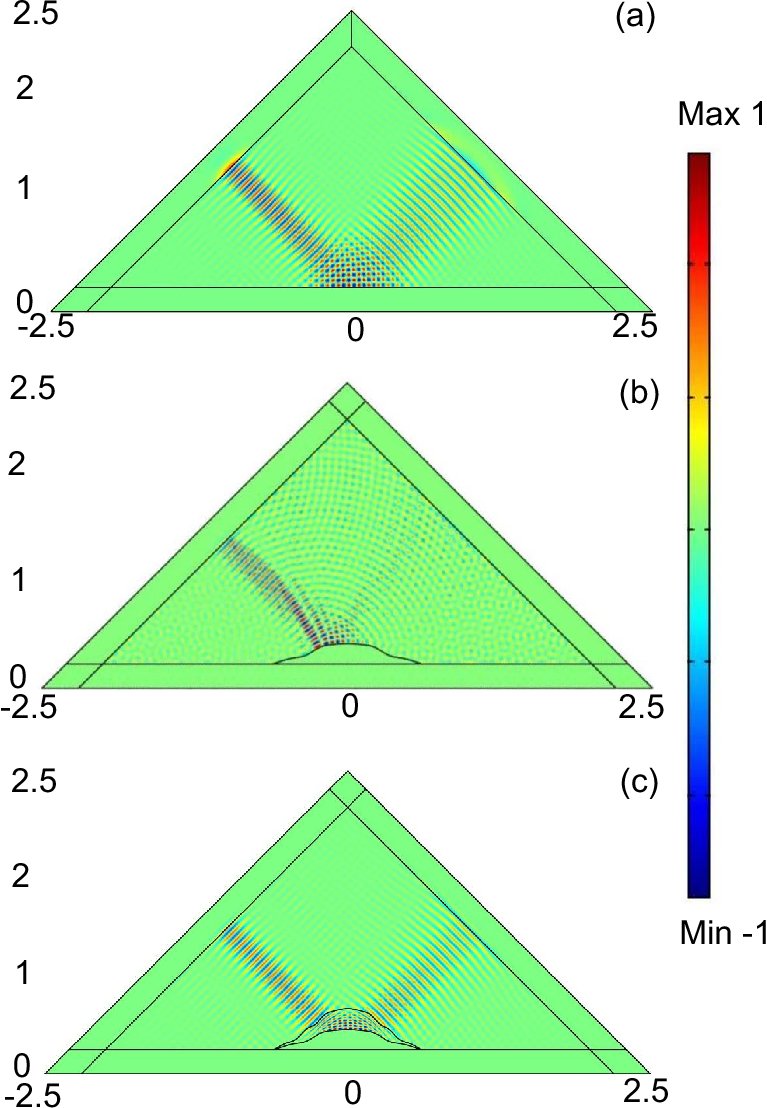}}
\caption{Diffraction by a Gaussian beam under oblique incidence at
wavelength $\lambda=0.06$: we set $H_3^i =
\exp(-1/2{(1/\sqrt{2}(x+y))}^2/0.1^2)$ on the left side of the inner
triangle; 2D plot of the real part of the component $H_3$ of the
magnetic field.
(a) Flat mirror; (b) Deformed mirror; (c) Same as (b) with a
carpet.} \label{fig4}
\end{figure}

We then repeat the same simulation with a Gaussian beam in order to
further analyse the effect of the carpet in a more realistic
physical situation. We report these computations in Fig. \ref{fig4}
where it should be noted that the beam reflected by the carpet
appears to have a waist closer to that of the incident beam than in
the case of a flat mirror. This might be attributed to the fact that
the optical path followed by the center of the beam is smaller in
the case of a carpet.

Let us finally consider the linear geometric transform:
\begin{equation}
\left\{
\begin{array}{ll}
x' &= x(r,\theta) \; , 0<r<\rho(\theta) \; , \; 0<\theta<2\pi \; , \\
y' &= y(r,\theta) \; , 0<r<\rho(\theta) \; , \; 0<\theta<2\pi \; , \\
z' &= \frac{z_2(x,y)-z_1(x,y)}{z_2(x,y)}z + z_1(x,y) \; , \; 0< z
<z_2(x,y) \; ,
\end{array}
\right. \label{transfo3D}
\end{equation}
where $z'$ is a stretched vertical coordinate. It is easily seen
that this linear geometric transform maps the arbitrary domain
$D=\bigcup_{(u,v)\in{(0,1)}^2}\{(x(u,v),y(u,v))\}$ within the plane
$xy$ onto the surface $z'=z_1(x,y)$, and leaves the surface
$z=z_2(x,y)$ unchanged. Importantly, there is a one-to-one
correspondence between the domain $D$ and the surfaces $z'=z_1$ and
$z'=z_2$. The surfaces $z_1$ and $z_2$ are assumed to be
differentiable, and this ensures that the carpet won't display any
singularity on its inner boundary.


The linear transform (\ref{transfo3D}) is expressed in a Cartesian
basis as: ${\bf J}_{xx'}=\left(
\begin{array}{ccc}
                1 & 0 & 0\\
             0 & 1 & 0\\
             \frac{\partial z}{\partial x'} & \frac{\partial z}{\partial y'} & \frac{1}{\alpha}\\
               \end{array}\right)$
where
$\alpha=(z_2-z_1)/z_2$ and from the chain rule
\begin{equation}
\begin{array}{ll}
&\displaystyle{\frac{\partial z}{\partial x'}= z_2
\frac{z'-z_2}{{(z_2-z_1)}^2} \frac{\partial z_1}{\partial x}- z_1
\frac{z'-z_1}{{(z_2-z_1)}^2}
\frac{\partial z_2}{\partial x}} \; , \\
&\displaystyle{\frac{\partial z}{\partial y'}= z_2
\frac{z'-z_2}{{(z_2-z_1)}^2} \frac{\partial z_1}{\partial y}- z_1
\frac{z'-z_1}{{(z_2-z_1)}^2} \frac{\partial z_2}{\partial y}}\; .
\end{array}
\end{equation}

\begin{figure}[h!]
\centerline{ \vspace{0cm}
\includegraphics[width=8cm,angle=0]{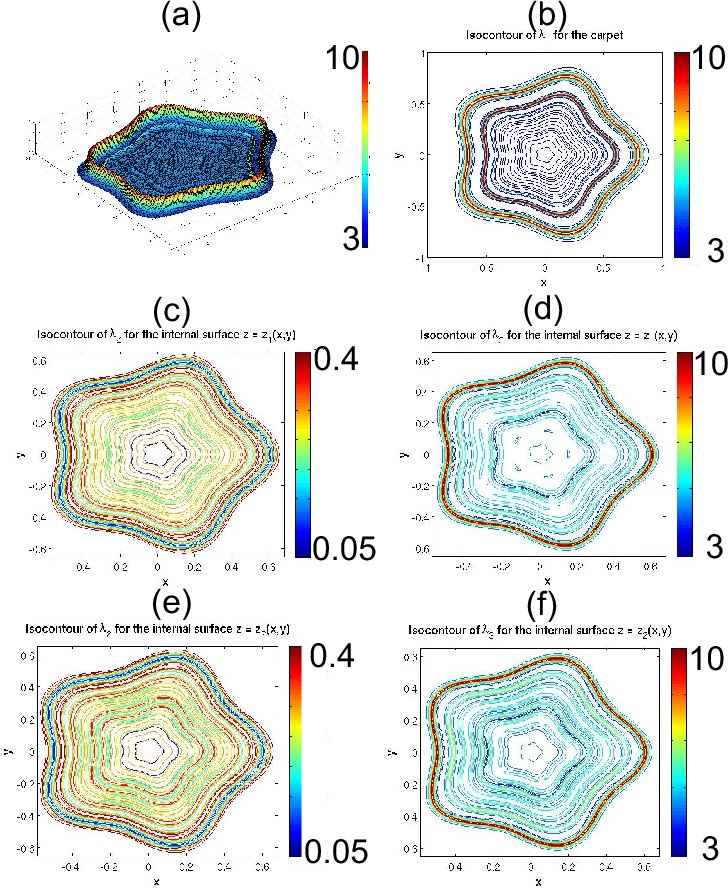}}
\vspace{1cm} \caption{(a) 3D plot of $\lambda_1$; (b) Corresponding
iso-contours; (c-f) Iso-contours of the eigenvalues
$\lambda_i(x,y)$, $i=2,3$ of ${\bf T}^{-1}$ given by (\ref{invt3d})
on the inner boundary $z_1(x,y)$ (b,c,d) and the outer boundary
$z_2(x,y)$ (b,e,f) of the carpet; We note that iso-contours of
$\lambda_1$ are the same whatever the altitude $z$.} \label{fig6}
\end{figure}

This leads to the inverse symmetric tensor ${\bf T}^{-1}$ which is
fully described by seven non vanishing entries in a Cartesian basis:
\begin{equation}
\begin{array}{ll}
(T^{-1})_{11}=(T^{-1})_{22}=\displaystyle \frac{1}{\alpha},
(T^{(-1)})_{13}=(T^{-1})_{31}=-\displaystyle \frac{\partial
z}{\partial x'},\\
(T^{-1})_{23}=(T^{-1})_{32}=-\displaystyle \frac{\partial
z}{\partial y'},\\
(T^{-1})_{33}=\displaystyle \left( 1+{\left(\frac{\partial
z}{\partial x'}\right)}^2+{\left(\frac{\partial z}{\partial
y'}\right)}^2\right) \alpha \; .
\end{array}
\label{invt3d}
\end{equation}
We note that the entries of the transformation matrix in
(\ref{invt3d}) reduce to those of (\ref{invt2d}) when
$\frac{\partial z}{\partial x'}$ vanishes. The corresponding
eigenvalues have the similar structure to (\ref{lambda2d}) and are
once again strictly positive and bounded, see Fig. \ref{fig6}, hence
the material parameters are non-singular.

Let us now compute the total electromagnetic field for a plane wave
normally incident upon a three-dimensional carpet. We have
implemented the weak form of this scattering problem in the finite
element package COMSOL using second order finite edge elements which
behave nicely under geometric changes. Perfectly Matched Layers
(PMLs), which can be seen as a stretch of coordinates, further
enable us to model the unbounded domain. We choose  the electric
field ${\bf E}=(E_1,E_2,E_3)(x,y,z)$ as the unknown in the Hilbert
space $H({\rm curl},\Omega)=\{{\bf v}\in {[L^2(\Omega)]}^3\; , {\rm
curl}{\bf v}\in {[L^2(\Omega)]}^3\}$ of curl-conforming fields
\cite{nicolet}, and therefore look for solutions of
\begin{equation}
\nabla\times\left( \underline{\underline{\mu'}}^{-1}\nabla\times
{\bf E} \right) - k_0^2 \underline{\underline{\varepsilon'}}{\bf E}
= {\bf 0} \; , \label{weak}
\end{equation}
where $k_0=\omega\sqrt{\mu_0\varepsilon_0}=\omega/c$ is the
wavenumber, $c$ being the speed of light in vacuum, and
$\underline{\underline{\varepsilon'}}$ and
$\underline{\underline{\mu'}}$ are defined by Eqs. (\ref{epsmuT}).
Also, ${\bf E}={\bf E}_i+{\bf E}_d$, where ${\bf E}_i$ is the
incident field (here a field approximating a plane wave incident
from the top which is generated by a constant field on a flat
surface on the upper part of the computational domain) and ${\bf
E}_d$ is the diffracted field which decreases inside the PMLs. We
note that we also used this setting to retrieve our former
computations assuming an electric field with the form $(E_1,E_2,0)$
in (\ref{weak}) to take advantage of pull-back properties of
edge-elements, leading again to Fig. \ref{fig1} when we compute the
curl of the numerical solution and plot the real part of the
longitudinal component of $(0,0,H_3)$

In this three-dimensional setting, we consider a plane wave incident
from above at normal incidence: ${\bf E}_i=e^{-ikz}{\bf e}_3$, with
wavenumber $k=2\pi/0.3$. The carpet has inner and outer surfaces
given by:
\begin{equation}
\begin{array}{ll}
&z_i(x,y)=h_i \left(
e^{-\frac{1}{2}{(\frac{\rho(\theta)}{\sigma})}^2} -\frac{1}{8}
\right) \\
&+c_i \sin\left(d_i.h_i \left(
e^{-\frac{1}{2}{(\frac{\rho(\theta)}{\sigma})}^2} -\frac{1}{8}
\right) \right) \; , i=1,2 \; ,
\end{array}
\end{equation}
where $\rho(\theta)=r(1-0.1\cos(5\theta))$ with $r=\sqrt{x^2+y^2}$
and $\theta=2\arctan(y/(x+\sqrt{x^2+y^2}))$ with $h_1=0.2$,
$h_2=0.3$, $c_1=0.003$, $c_1=0.005$ $d_1=d_2=200$ and $\sigma=0.3$.

\begin{figure}[h!]
\centerline{
\includegraphics[width=8cm,angle=0]{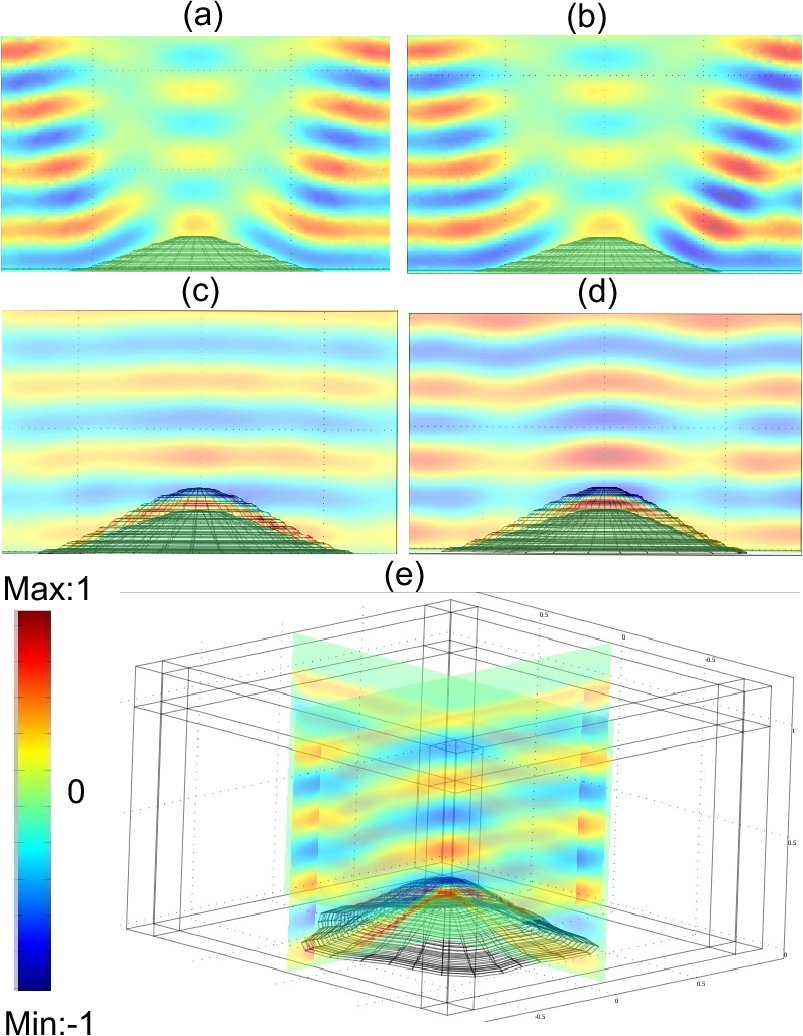}}
\vspace{0.3cm} \caption{Diffraction of a normally incident plane
wave by a deformed mirror surrounded by a 3D carpet at wavelength
$\lambda=0.3$; (a)-(d) 2D plots of the real part of the component
$E_3$ of the electric field in the planes $x0z$, $y0z$ for the bump
on its own (upper panel) and with a carpet (middle panel); (e) 3D
plot of the real part of $E_3$: Cartesian Perfectly Matched Layers
were implemented in the domains surrounding the central cubic region
(lower panel).} \label{fig2}
\end{figure}

It is clearly seen from panels (c) and (d) in Fig. \ref{fig2} that
although of a complex non-convex shape, see (e), the carpet reflects
the plane wave nearly like a flat ground plane would. When the bump
is not covered by the carpet, the scattering is much worse, see (a)
and (b).

\begin{figure}[h!]
\vspace{1.4cm} \centerline{
\includegraphics[width=8cm,angle=0]{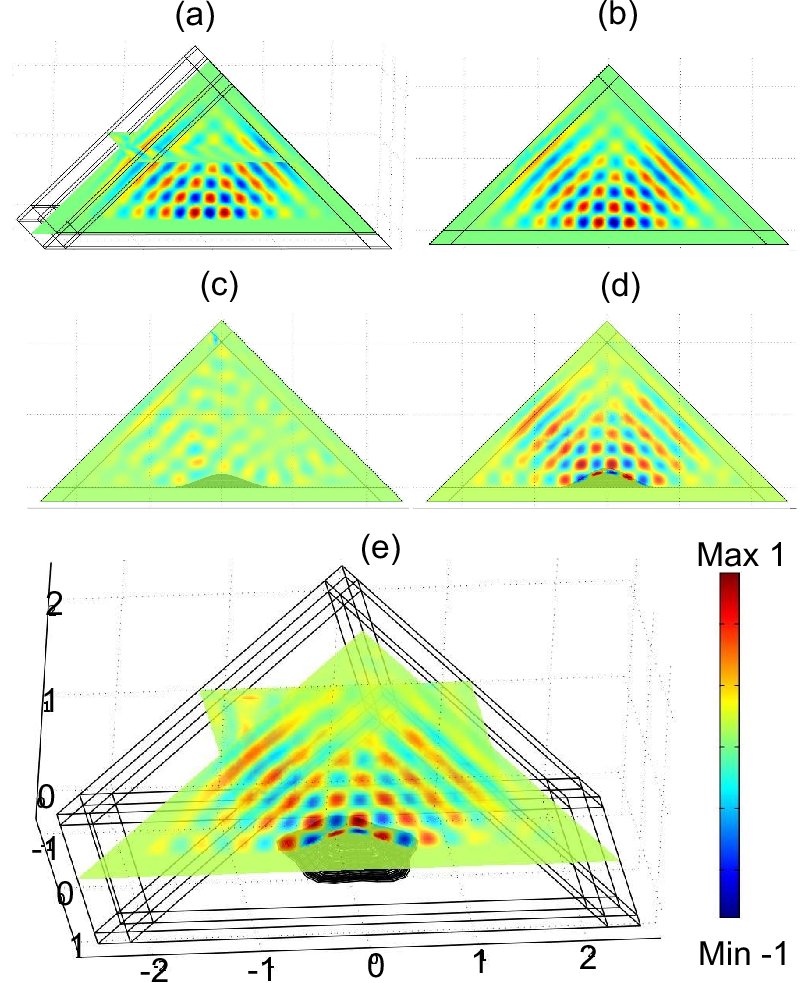}}
\vspace{1.4cm} \caption{Diffraction of a Gaussian beam in oblique
incidence at wavelength $\lambda=0.3$ (we set $E_3^i =
\exp(-1/2(z^2/0.4^2+{(1/\sqrt{2}(-x-y))}^2/0.4^2))$ on the left-hand
side of the inner tetrahedron) by a deformed mirror surrounded by a
3D carpet; (b)-(d) 2D plots of the real part of the component $E_3$
of the electric field in the plane $x0z$ for the flat mirror on its
own, ((b) upper right panel), the bump on its own ((c) middle left
panel) and with a carpet ((d) middle right panel); (a) and (e) 3D
plot of the real part of $E_3$ for a flat mirror (upper left panel)
and a deformed mirror surrounded by the carpet (lower panel);
Perfectly Matched Layers were implemented in the prism by applying a
rotation of $\pi/4$ radians in the coordinate axes.} \label{fig5}
\end{figure}

Finally, we repeat these simulations for a Gaussian beam in oblique
incidence (making an angle $\pi/4$ with the vertical axis). This
requires a computational domain shaped as a prism, see Fig.
\ref{fig5}. We note that the plots of the field are indeed symmetric
with the $xOz$ plane in the case of a flat mirror, see Fig.
\ref{fig5}(a-b), and a deformed mirror surrounded by the carpet, see
Fig. \ref{fig5}(d-e). However, the diffraction by a deformed mirror
is clearly giving rise to an asymmetric field, see Fig.
\ref{fig5}(c).

%

In this paper, we have shown that it is possible to design
two-dimensional and three-dimensional carpets of an arbitrary shape,
using a similar approach to Fourier-based cloaks \cite{nzg2008}.
Such carpets do not exhibit any singular material parameters on
their inner boundary, unlike invisibility cloaks, as they are based
upon a one-to-one geometric transform. The next step towards the
realization of such carpets might involve some structural elements
such as conducting thin-straight wires and split ring resonators
\cite{pendry_IEEE} to tune the permittivity and permeability to
required values depending upon light polarization. The rapid
experimental progress in the construction of carpets getting close
to optical frequencies
\cite{smithcarpet,zhangcarpet1,gabrielli-carpet} suggests that our
designs might soon come to life.


The authors acknowledge insightful discussions with A. Diatta, G.
Demesy, M. Farhat, A. Nicolet and F. Zolla.

\end{document}